\documentclass [6pt,a4paper]{article}
\usepackage{bbm}
\usepackage{amsfonts}
\usepackage{mathrsfs}
\usepackage {amssymb}
\usepackage {amsmath}
\usepackage{amsthm}
\usepackage{latexsym}
\def\dse#1{\vskip 0.6cm\noindent
        {\large\bf #1}
        \vskip 0.4cm}

\def\dse#1{\vskip 0.6cm\noindent
        {\large\bf #1}
        \vskip 0.4cm}
\usepackage{lastpage}
\usepackage{epsfig}

\begin{document}
\begin{center}
\textbf{\large{Repeated-root constacyclic codes of length $3lp^{s}$ and their dual codes }}\footnote { E-mail
addresses:
 liuli-1128@163.com(Li Liu), lilanqiang716@126.com(Lanqiang Li).\\
This research is supported by the National Natural Science
Foundation of China (No.61572168,No.11401154) and the Anhui provincial Natural Science Foundation under Grant (No.158085MA13).}
\end{center}

\begin{center}
{ {Li Liu, \  Lanqiang Li, \ Xiaoshan Kai, \ Shixin Zhu} }
\end{center}

\begin{center}
\textit{\footnotesize School of Mathematics, Hefei University of
Technology, Hefei 230009, Anhui, P.R.China }
\end{center}

\noindent\textbf{Abstract:} Let $p\neq3$ be any prime and $l\neq3$ be any odd prime with gcd$(p,l)=1$. The multiplicative group $F_{q}^{*}=\langle\xi\rangle$ can be decomposed into mutually disjoint union of gcd$(q-1,3lp^{s})$ cosets over the subgroup $\langle\xi^{3lp^{s}}\rangle$, where $\xi$ is a primitive $(q-1)$th root of unity. We classify all repeated-root constacyclic codes of length $3lp^{s}$ over the finite field $F_{q}$  into some equivalence classes by this decomposition, where $q=p^{m}$, $s$ and $m$ are positive integers. According to these equivalence classes, we explicitly determine the generator polynomials of all repeated-root constacyclic codes of length $3lp^{s}$ over $F_{q}$ and their dual codes. Self-dual cyclic codes of length $3lp^{s}$ over $F_{q}$ exist only when $p=2$. And we give all self-dual cyclic codes of length $3\cdot2^{s}l$ over $F_{2^{m}}$ and their enumeration.\\

\noindent\emph{Keywords}: Repeated-root constacyclic codes, Cyclic codes, Dual codes, Generator polynomial

\dse{1~~Introduction}
Constacyclic codes over finite fields play a very important role in the theory of error-correcting codes. More importantly, constacyclic codes have practical applications. As these codes have rich algebraic structures, they can be efficiently encoded and decoded using shift registers. They also have very good error-correcting properties. All of those explain their preferred role in engineering.

Repeated-root cyclic codes were first investigated in the $1990s$ by Castagnoli in $[1]$ and Van Lint in $[2]$, where it was proved that repeated-root cyclic codes have a concatenated construction, and are asymptotically bad. However, it is well known that there still exist a few optimal such codes (see $[12-14]$), which encourages many scholars to study the class of codes. For example, Dinh determined the generator polynomials of all constacyclic codes and their dual codes over $F_{q}$ of length $2p^{s}, 3p^{s}$ and $6p^{s}$ in $[3-5]$. Since then, these results have been extended to more general code lengths. In $2012$, Bakshi and Raka gave the generator polynomials of all constacyclic codes of length $2^{t}p^{s}$ over $F_{q}$ in $[6]$, where $q$ is a power of an odd prime $p$. In $2014$, Chen et.al studied all constacyclic codes of length $lp^{s}$ over $F_{q}$ in $[7]$, where $l$ is a prime different from $p$. In $[7]$, all constacyclic codes of length $lp^{s}$ over $F_{q}$ and their dual codes were obtained, and all self-dual and all linear complementary dual constacyclic codes were given. Recently, in $[8]$, Sharma explicitly determined the generator polynomials of all repeated-root constacyclic codes of length $l^{t}p^{s}$ over $F_{p^{m}}$ and their dual codes. Further, they listed all self-dual cyclic and negacyclic codes and also determined all self-orthogonal cyclic and negacyclic codes of length $l^{t}p^{s}$ over $F_{p^{m}}$. What's more, Chen et.al studied all constacyclic codes of length $2l^{m}p^{s}$ over $F_{q}$ of characteristic $p$ in $[9]$, and they gave the characterization and enumeration of all linear complementary dual and self-dual constacyclic codes of  length $2l^{m}p^{s}$ over $F_{q}$. In the conclusion of their paper, they proposed to study all constacyclic codes of length $kl^{m}p^{s}$ over $F_{q}$, where $p$ is the characteristic of $F_{q}$, $l$ is an odd prime different from $p$, and $k$ is a prime different from $l$ and $p$. However, this is not an easy work.

In this paper, we study all constacyclic codes of length $3lp^{s}$ over $F_{q}$, where $p\neq3$ is  any prime and $l\neq3$ is any odd prime with $\textrm{gcd}(p,l)=1$. The article is organized as follows. In section 3, we decompose the multiplicative cyclic group $F_{q}^{*}=\langle\xi\rangle$ into mutually disjoint union of cosets of $\langle\xi^{3lp^{s}}\rangle$, which are one-to-one correspondence to the equivalence classes of all constacyclic. Based on this decomposition, Section 4 explicitly determines the generator polynomials of all $\lambda-$constacyclic codes of length $3lp^{s}$ over $F_{q}$ and their dual codes, where $\lambda$ is any none-zero element of $F_{q}$ and $q=p^{m}$ is a power of prime. As an application, we give all self-dual cyclic codes of length $3\cdot2^{s}l$ over $F_{2^{m}}$ and their enumeration in Section $5$.

\dse{2~~Preliminaries}
  Let $F_{q}$ be the finite field of order $q=p^{m}$, where $p\neq3$ is a prime and the characteristic of the field, and $m$ is a positive integer. Let $F_{q}^{*} =\langle\xi\rangle$ be the multiplicative cyclic group of none-zero elements of $F_{q}$, where $\xi$ is a primitive $(q-1)$th root of unity.

  For any element $\lambda\in F_{q}^{*}$, $\lambda-$constacyclic codes of length $n$ over $F_{q}$ are regarded as the ideals $\langle g(x)\rangle$ of the quotient ring $F_{q}[x]/(x^{n}-\lambda)$, where $g(x)|x^{n}-\lambda$. Further, the definition of the dual code of code $C$ is as follows:
 $$C^{\perp}=\{x\in F_{q}^{n}|x\cdot y=0, \forall y\in C\},$$
 where $x\cdot y$ denotes the Euclidean inner product of $x$ and $y$ in $F_{q}^{n}$. The code $C$ is called to be a self-orthogonal code if $C\subseteq C^{\perp}$ and a self-dual code if $C=C^{\perp}$. Let $C$ be a $\lambda-$constacyclic code of length $n$ over $F_{q}$ generated by a polynomial $g(x)$,
  i.e., $C=\langle g(x)\rangle$. As $g(x)|x^{n}-\lambda$, then there must be a polynomial $h(x)\in F_{q}[x]$ such that $h(x)=\frac{x^{n}-\lambda}{g(x)}$. It is clear that $h(x)$ is also monic if $g(x)$ is monic. The polynomial $h(x)$ is called the parity check polynomial of code $C$. It is well known that the dual code $C^{\perp}$ is generated by $h(x)^{*}$, where $h(x)^{*}$ is the reciprocal polynomial of $h(x)$. For any $f(x)\in F_{q}[x]$, the reciprocal polynomial of $f(x)$ is defined as $f(x)^{*}=f(0)^{-1}x^{\textrm{deg}(f(x))}f(\frac{1}{x})$. Obviously, $(f_{1}f_{2})^{*}=f_{1}^{*}f_{2}^{*}$, and $(f_{1}^{*})^{*}=f_{1}$, for any polynomials $f_{1}(x),f_{2}(x)\in F_{q}[x]$.

  Let $n$ be any positive integer. For any integer $s$, $0\leq s\leq n-1$, we have the definition of $q-$cyclotomic coset of $s$ modulo $n$ is as follows:
$$C_{s}=\{s,sq,...,sq^{n_{s}-1}\},$$
where $n_{s}$ is the least positive integer such that $sq^{n_{s}}\equiv s (\textrm{mod}~n)$. Then it is easy to see that $n_{s}$ is equal to the multiplicative order of $q$ modulo $\frac{n}{\textrm{gcd}(s,n)}$. If $\alpha$ denotes a primitive $n$th root of unity in some extension field of $F_{q}$, then the polynomial $M_{s}(x)=\prod_{i\in C_{s}}(x-\alpha^{i})$ is the minimal polynomial of $\alpha^{s}$ over $F_{q}$, and $$x^{n}-1=\prod M_{s}(x)$$ gives the factorization of $(x^{n}-1)$ into irreducible factors over $F_{q}$, where $s$ runs over a complete set of representatives from distinct $q-$cyclotomic coset modulo $n$.

Obviously, when $n=l$, where $l\neq3$ is an odd prime with gcd$(l,p)=1$, we get that all the distinct $q-$cyclotomic cosets modulo $l$ are $C_{0}=\{0\}$ and $C_{k}=\{g^{k},g^{k}q,...,g^{k}q^{n_{k}-1}\},$ for any integer $k$, $1\leq k\leq e=\frac{\phi(l)}{f}$, by $[15,\textrm{Theorem}~1]$, where $g$ is a fixed generator of the cyclic group $Z_{l}^{*}$, $f=\textrm{\textrm{ord}}_{l}(q)$ is the multiplicative order of $q$ in $Z_{l}^{*}$, and $\phi$ is Euler's phi-function. Therefore, the irreducible factorization of $x^{l}-1$ over $F_{q}$ is given by $$x^{l}-1=M_{0}(x)M_{1}(x)M_{2}(x)...M_{e}(x),$$ where $M_{i}(x)=\prod_{j\in C_{i}}(x-\eta^{i})$ with $\eta$ being a primitive $l$th root of unity.

In addition, we determine all the distinct $q^{3}-$cyclotomic cosets modulo $l$, which is needed to prove our main results. There exist two subcases. When $\textrm{gcd}(f,3)=1$, it is easy to prove that $f=\textrm{ord}_{l}(q)=\textrm{ord}_{l}(q^{3})$, then $\langle q\rangle=\langle q^{3}\rangle$ in $Z_{l}^{*}$. According to the definition of $q^{3}-$cyclotomic coset modulo $l$, we have that $C_{0}$ and $C_{k}$, $1\leq k\leq e=\frac{\phi(l)}{f}$, are also all of the distinct $q^{3}-$cyclotomic cosets modulo $l$. When $\textrm{gcd}(f,3)=3$, we prove that $\textrm{ord}_{l}(q^{3})=\frac{f}{3}$. It is easy to verify that $A_{0}=\{0\}, $ $$A_{k}=\{g^{k},g^{k}q^{3},...,g^{k}q^{3(\frac{f}{3}-1)}\}, $$ $$A_{kq}=\{g^{k}q,g^{k}qq^{3},...,g^{k}qq^{3(\frac{f}{3}-1)}\},$$ $$A_{kq^{2}}=\{g^{k}q^{2},g^{k}q^{2}q^{3},...,g^{k}q^{2}q^{3(\frac{f}{3}-1)}\},$$  consist of all the distinct $q^{3}-$cyclotomic cosets modulo $l$, where $1\leq k\leq e$. Then we have the irreducible factorization of $x^{l}-1$ in $F_{q^{3}}[x]$ as follows: $$x^{l}-1=A_{0}(x)A_{1}(x)A_{q}(x)A_{q^{2}}(x)A_{2}(x)A_{2q}(x)A_{2q^{2}}(x)...A_{e}(x)A_{eq}(x)A_{eq^{2}}(x),$$
where $A_{0}(x)=(x-1)$, $A_{k}(x)=\prod_{s\in A_{k}}(x-\eta^{s})$, $A_{kq}(x)=\prod_{t\in A_{kq}}(x-\eta^{t})$ and $A_{kq^{2}}(x)=\prod_{j\in A_{kq^{2}}}(x-\eta^{j})$, $1\leq k\leq e$.

 We also give all the distinct $q-$cyclotomic cosets modulo $3l$. As gcd$(q,3)=1$, we have $q^{\phi(3)}\equiv 1(\textrm{mod}~3)$ by Euler's Theorem, i.e., $q^{2}\equiv 1(\textrm{mod}~3)$. Then, it is simple to verify that
       \begin{eqnarray*}
\textrm{ord}_{3l}(q)=
\left\{ {{\begin{array}{ll}
 {f}, & {q\equiv 1(\textrm{mod}~3)};\\
 {f}, & {q\equiv 2(\textrm{mod}~3) ~\textrm{with}~ f ~\textrm{even}}; \\
 {2f}, & {q\equiv 2(\textrm{mod}~3) ~\textrm{with} ~f~ \textrm{odd}}. \\
\end{array} }} \right .
\end{eqnarray*}
From $[16,\textrm{Chapter}~8]$, there exists a primitive root $r$ modulo $l$ such that $\textrm{gcd}(\frac{r^{l-1}-1}{l},l)=1$. Assume that $g=r+(1-r)l^{2}$, we have $g^{l-1}-1\equiv (r+(1-r)l^{2})^{l-1}-1\equiv r^{l-1}-1(\textrm{mod}~l^{2})$. Therefore, $\textrm{gcd}(\frac{g^{l-1}-1}{l},l)=\textrm{gcd}(\frac{r^{l-1}-1}{l},l)=1$. It's clear that $g$ is a primitive root modulo $l^{t}$, $t\geq1$, such that $g\equiv 1(\textrm{mod}~3)$.

We give all the distinct $q-$cyclotomic cosets modulo $3l$ by the following lemma.\\

\noindent\textbf{Lemma 2.1.} (I) If $q\equiv 1(\textrm{mod}~3)$, then we have that all the distinct $q-$cyclotomic cosets modulo $3l$ are given by $$B_{0}=\{0\},B_{l}=\{l\},B_{-l}=\{-l\},$$
$$B_{ag^{k}}=\{ag^{k},ag^{k}q,...,ag^{k}q^{f-1}\},$$ for $a\in R=\{1,-1,3\}$ and $0\leq k\leq e-1$.
\\(II) If $q\equiv 2(\textrm{mod}~3)$ and $f$ is even, we have that all the distinct $q-$cyclotomic cosets modulo $3l$ are given by $B_{0}=\{0\},B_{l}=\{l,lq\},$
$$B_{g^{k^{'}}}=\{g^{k^{'}},g^{k^{'}}q,...,g^{k^{'}}q^{f-1}\}, ~\textrm{for} ~0\leq k^{'}\leq 2e-1,$$
$$B_{3g^{k}}=\{3g^{k},3g^{k}q,...,3g^{k}q^{f-1}\},~\textrm{for}~ 0\leq k\leq e-1.$$
 \\(III) If $q\equiv 2(\textrm{mod}~3)$ and $f$ is odd, we have that all the distinct $q-$cyclotomic cosets modulo $3l$ are given by $$B_{0}=\{0\},B_{l}=\{l,lq\},$$
$$B_{g^{k}}=\{g^{k},g^{k}q,...,g^{k}q^{2f-1}\},$$
$$B_{3g^{k}}=\{3g^{k},3g^{k}q,...,3g^{k}q^{f-1}\},$$ for $0\leq k\leq e-1$.\\

\noindent\textbf{Proof.} (I) Firstly, we prove that the cyclotomic cosets $B_{ag^{k}}, 0\leq k\leq e-1$, are distinct. If there exist some $k_{1},k_{2}$, $0\leq k_{1},k_{2}\leq e-1$, such that $B_{ag^{k_{1}}}=B_{ag^{k_{2}}}$, then we have
 $$a_{1}g^{k_{1}}\equiv a_{2}g^{k_{2}}q^{j}(\textrm{mod}~3l),$$ for some integer j, where $a_{1},a_{2}\in R=\{1,-1,3\}.$ Therefore, we get $$\textrm{gcd}(a_{1}g^{k_{1}},3l)=\textrm{gcd}(a_{2}g^{k_{2}}q^{j},3l)=\textrm{gcd}(a_{2}g^{k_{2}},3l),$$ as $\textrm{gcd}(q,3l)=1$. From this, we can deduce $a_{1}=a_{2}$ or $a_{1}=-a_{2}=\pm 1$.

 If $a_{1}=-a_{2}=\pm 1$, then $$ -g^{k_{1}}\equiv g^{k_{2}}q^{j}(\textrm{mod}~3l), ~~  \textrm{i.e.,} ~~  -1\equiv g^{k_{1}-k_{2}}q^{j}(\textrm{mod}~3l),$$for some integer $j$. Due to $g\equiv 1(\textrm{mod}~3)$ and $q\equiv1(\textrm{mod}~3)$, we deduce $-1\equiv1(\textrm{mod}~3)$. This is a contradiction.

 If $a_{1}=a_{2}$, assume that $a_{1}=a_{2}=a$, then we have $$ag^{k_{1}}\equiv ag^{k_{2}}q^{j}(\textrm{mod}~3l), ~~ \textrm{i.e.,}  ~~ g^{k_{1}-k_{2}}\equiv q^{j}(\textrm{mod}~l),$$for some integer $j$. Further, we have $g^{(k_{1}-k_{2})f}\equiv q^{jf}\equiv1(\textrm{mod}~l)$. As $g$ is a primitive root modulo $l$, we get $\phi(l)|(k_{1}-k_{2})f$, i.e., $\textrm{e}=\frac{\phi(l)}{f}|k_{1}-k_{2}$. Since $0\leq k_{1},k_{2}\leq e-1$, we must have $k_{1}=k_{2}$. Secondly, we get
 \begin{eqnarray*}
        |B_{0}|+|B_{l}|+|B_{-l}|+\sum_{a\in R}\sum_{k=0}^{e-1}|B_{ag^{k}}|    &=&3+\sum_{a\in R}\sum_{k=0}^{e-1}f\\
              &=&3+\sum_{a\in R}ef\\
              &=&3+3\phi(l)\\
              &=&3l.
\end{eqnarray*}
So, the conclusion (I) holds. The conclusions (II) and (III) are also established in a similar way.\qed\\

Assume that $B_{o}(x),B_{l}(x),B_{-l}(x)$ and $B_{ag^{k}}(x)$ are the minimal polynomials of the corresponding cosets $B_{o},B_{l},B_{-l}$ and $B_{ag^{k}}$. From the above lemma, we get the following theorem immediately.\\

\noindent\textbf{Theory 2.2.} The irreducible factorization of $x^{3l}-1$ over $F_{q}$ is given as follows:\\(I) If $q\equiv1(\textrm{mod}~3)$, then\\$$x^{3l}-1=B_{0}(x)B_{l}(x)B_{-l}(x)\prod_{a\in R} \prod_{k=0}^{e-1}B_{ag^{k}}(x),$$ where $a\in R=\{1,-1,3\}$ and $0\leq k\leq e-1$.\\
(II) If $q\equiv2(\textrm{mod}~3)$ and $f$ is even, then
$$x^{3l}-1=B_{0}(x)B_{l}(x)\prod_{k^{'}=0}^{2e-1}B_{g^{k^{'}}}(x)\prod_{k=0}^{e-1}B_{3g^{k}}(x),$$
 where $0\leq k\leq e-1,0\leq k^{'}\leq 2e-1$.\\
(III) If $q\equiv2(\textrm{mod}~3)$ and $f$ is odd, then
$$x^{3l}-1=B_{0}(x)B_{l}(x)\prod_{k=0}^{e-1}B_{g^{k}}(x)B_{3g^{k}}(x),$$
 where $0\leq k\leq e-1$.\\

The next two lemmas give the necessary and sufficient conditions for judging the reducibility of binomials and trinomial, which were given by Wan in $[17]$.\\

\noindent\textbf{Lemma 2.3.} Suppose that $n\geq2$, Let $k=\textrm{ord}(a)$ be the multiplicative order of $a$, for any $a\in F_{q}^{*}$. Then, the binomial $x^{n}-a$ is irreducible over $F_{q}$ if and only if \\(i) Every prime divisor of $n$ divides $k$, but not $\frac{(q-1)}{k} $;
\\(ii) If $4|n$, then $4|(q-1)$.\\

\noindent\textbf{Lemma 2.4.} Let $t$ be a positive integer, and $H(x)\in F_{q}[x]$ be irreducible over $F_{q}$ with $\textrm{deg}(H(x))=n$, $x$ does not divide $H(x)$, and $e$ denotes the order of any root of $H(x)$. Then $H(x^{t})$ is irreducible over $F_{q}$ if and only if \\(i) Each prime divisor of $t$ divides $e$;\\(ii) $\textrm{gcd}(t,\frac{q^{n}-1}{e})=1$;\\
(iii) If $4|t$, then $4|(q^{n}-1)$.

\dse{3~~ A classification of constacyclic codes of length $3lp^{s}$ }
Let $\xi$ be a primitive $(q-1)$th root of unity, and $F_{q}^{*}=\langle \xi\rangle$ be a cyclic group of order $(q-1)$ as before. It is easy to verify that $\langle \xi^{3lp^{s}}\rangle=\langle\xi^{3l}\rangle=\langle\xi^{d}\rangle$ and the index $|F_{q}^{*}:\langle \xi^{3lp^{s}}\rangle|=d$, where $d=\textrm{gcd}(q-1,3lp^{s})$. Thus, the multiplicative cyclic group $F_{q}^{*}$ can be decomposed into mutually disjoint union of cosets over the subgroup $\langle\xi^{3lp^{s}}\rangle$ as follows:\\

\noindent\textbf{Lemma 3.1.}
$F_{q}^{*}=\langle\xi\rangle=\langle\xi^{d}\rangle\bigcup\xi^{p^{s}}\langle\xi^{d}\rangle\bigcup...\bigcup\xi^{p^{s}(d-1)}\langle\xi^{d}\rangle$,
where $d=\textrm{gcd}(q-1,3lp^{s})$.\\

 According to the properties of the coset, we obtain the following lemma immediately. \\

 \noindent\textbf{Lemma 3.2.} For any two none-zero elements $\lambda$ and $\mu$ of $F_{q}$, there exists some integer $j$, $0\leq j\leq d-1$ such that $\lambda,\mu\in\xi^{jp^{s}}\langle\xi^{d}\rangle$ if and only if $\lambda^{-1}\mu\in\langle\xi^{d}\rangle$, where $d=\textrm{gcd}(q-1,3lp^{s})$.\\

 If $\lambda$ and $\mu$ are in the same coset, we build a one-to-one correspondence between $\lambda-$constacyclic codes and $\mu-$constacyclic codes of length $3lp^{s}$ over $F_{q}$ as following theorem, which shows that  $\lambda-$constacyclic codes and
$\mu-$constacyclic codes are equivalent.\\

 \noindent\textbf{Theorem 3.3.} Let $\lambda,\mu\in F_{q}^{*}$. Then there exists some integer $a\in F_{q}^{*}$ such that
 \begin{eqnarray*}
&&\varphi:F_{q}[x]/(x^{3lp^{s}}- \mu)\rightarrow F_{q}[x]/(x^{3lp^{s}}-\lambda),\\
&&f(x)\mapsto f(ax),
\end{eqnarray*}
is an isomorphism if and only if $\lambda,\mu\in\xi^{jp^{s}}\langle\xi^{d}\rangle$, where $0\leq j\leq d-1$. \\

\noindent\textbf{Proof.} $ (\Rightarrow) $ If $\varphi$ is an isomorphism, then we have
 $$\mu=\varphi(\mu)=\varphi(x^{3lp^{s}})=(\varphi(x))^{3lp^{s}}=(ax)^{3lp^{s}}=a^{3lp^{s}}x^{3lp^{s}}=a^{3lp^{s}}\lambda,$$
i.e., $\lambda^{-1}\mu=a^{3lp^{s}}$. As $a=\xi^{k}\in F_{q}^{*}$, for some positive integer $k$, then  $\lambda^{-1}\mu=\xi^{k\cdot3lp^{s}}\in \langle\xi^{d}\rangle$. By Lemma 3.2, we get that there exists $j$, $0\leq j\leq d-1$, such that $\lambda,\mu\in\xi^{jp^{s}}\langle\xi^{d}\rangle$.

$(\Leftarrow)$ If there exists $j$, $0\leq j\leq d-1$, such that $\lambda,\mu \in\xi^{jp^{s}}\langle\xi^{d}\rangle$, then we have $\lambda^{-1}\mu\in \langle\xi^{d}\rangle=\langle\xi^{3lp^{s}}\rangle$, by Lemma 3.2 again. Thus, $\lambda^{-1}\mu=\xi^{k\cdot3lp^{s}}$, for some integer $k$. Set $a=\xi^{k}$, then $\lambda a^{3lp^{s}}=\mu $. Further, it is easy to prove that the following map is an isomorphism:
\begin{eqnarray*}
&&\varphi:F_{q}[x]/(x^{3lp^{s}}-\mu )\rightarrow F_{q}[x]/(x^{3lp^{s}}-\lambda),\\
&&f(x)\mapsto f(ax).
\end{eqnarray*}\qed\\

From Theorem 3.3, we get the following obvious corollaries. \\

\noindent\textbf{Corollary 3.4.} For any two elements $\lambda$ and $\mu$ of $F_{q}^{*}$, a $\lambda-$constacyclic code is equivalent to an $\mu-$constacyclic code if and only if there exists $j$, $0\leq j\leq d-1$, such that $\lambda,\mu\in\xi^{jp^{s}}\langle\xi^{d}\rangle$. Further, a $\lambda-$constacyclic code and an $\mu-$constacyclic code are both equivalent to a $\xi^{jp^{s}}-$constacyclic code.\\

\noindent\textbf{Corollary 3.5.} Let  $\lambda$ be any element of $F_{q}^{*}$, then there exists some integer $j$, $0\leq j\leq d-1$, such that a $\lambda-$constacyclic code is equivalent to a $\xi^{jp^{s}}-$constacyclic code.\\

Obviously,  the Theorem 3.3 and its two corollaries show that all constacyclic codes of length $3lp^{s}$ over $F_{q}$ are classified into $d=\textrm{gcd}(q-1,3lp^{s})$ mutually disjoint classes. It is enough to consider $\lambda-$constacyclic codes, where $\lambda=\xi^{jp^{s}}$, $0\leq j\leq d-1$, and $d=\textrm{gcd}(q-1,3lp^{s})$, if we want to  determine all constacyclic codes of length $3lp^{s}$ over $F_{q}$. Therefore, we mainly study $\lambda-$constacyclic codes in Section 4.
\dse{4~~All constacyclic codes of length $3lp^{s}$ over $F_{q}$}
Let $f(x)$ be any polynomial of $F_{q}[x]$ and leading coefficient $a_{n}\neq 0$, we denote $\widehat{f}(x)=a_{n}^{-1}f(x)$. Then, $\widehat{f}(x)$ is said to be the monic polynomial of $f(x)$.

From the above discussion in the section 3, we know that the number of equivalence constacyclic classes is equal to $d=\textrm{gcd}(q-1,3lp^{s})$. It is obvious that the value of $d$ has the following four cases:
\\(i) $d=\textrm{gcd}(q-1,3lp^{s})=1$.\\(ii) $d=\textrm{gcd}(q-1,3lp^{s})=3$.\\(iii) $d=\textrm{gcd}(q-1,3lp^{s})=l$.\\(iv) $d=\textrm{gcd}(q-1,3lp^{s})=3l$.

\dse{4.1~~All constacyclic codes of length $3lp^{s}$ over $F_{q}$ when $d=1$}
From Lemma 3.1, we see that $F_{q}^{*}=\langle\xi\rangle$ is the decomposition of coset over the subgroup $\langle\xi^{d}\rangle$, when $d=\textrm{gcd}(q-1,3lp^{s})=1$. In this case, it is clear that all constacyclic codes of length $3lp^{s}$ over $F_{q}$ are equivalent to the cyclic codes. Therefore, we have the following theorem.\\

\noindent\textbf{Theorem 4.1.} Let $d=\textrm{gcd}(q-1,3lp^{s})=1$, then $\lambda-$constacyclic codes $C$ of length $3lp^{s}$ over $F_{q}$ are equivalent to the cyclic codes, for any $\lambda\in F_{q}^{*}$, i.e., there exists a unique element $a\in F_{q}^{*}$ such that $a^{3lp^{s}}\lambda=1$. Further, we have the map
 \begin{eqnarray*}
&&\varphi_{a}:F_{q}[x]/(x^{3lp^{s}}-1 )\rightarrow F_{q}[x]/(x^{3lp^{s}}-\lambda),\\
&&f(x)\mapsto f(ax),
\end{eqnarray*}
is an isomorphism, and the irreducible factorization of $x^{3lp^{s}}-\lambda$ in $F_{q}[x]$ is given by:\\
(I) If $f$ is even, then \\
$$x^{3lp^{s}}-\lambda=\widehat{B}_{0}(ax)^{p^{s}}\widehat{B}_{l}(ax)^{p^{s}}\prod_{k^{'}=0}^{2e-1}\widehat{B}_{g^{k^{'}}}(ax)^{p^{s}}\prod_{k=0}^{e-1}\widehat{B}_{3g^{k}}(ax)^{p^{s}},$$
 where $0\leq k\leq e-1,0\leq k^{'}\leq 2e-1$.

 Therefore, we have $$C=\langle\widehat{B}_{0}(ax)^{\varepsilon}\widehat{B}_{l}(ax)^{\rho}\prod_{k^{'}=0}^{2e-1}\widehat{B}_{g^{k^{'}}}(ax)^{\tau_{k^{'}}}\prod_{k=0}^{e-1}\widehat{B}_{3g^{k}}(ax)^{\upsilon_{k}}\rangle,$$
 $$C^{\perp}=\langle\widehat{B}_{0}(a^{-1}x)^{p^{s}-\varepsilon}\widehat{B}_{-l}(a^{-1}x)^{p^{s}-\rho}\prod_{k^{'}=0}^{2e-1}\widehat{B}_{-g^{k^{'}}}(a^{-1}x)^{p^{s}-\tau_{k^{'}}}\prod_{k=0}^{e-1}\widehat{B}_{-3g^{k}}(a^{-1}x)^{p^{s}-\upsilon_{k}}\rangle,$$\\
where $0\leq \varepsilon,\rho,\tau_{k^{'}},\upsilon_{k}\leq p^{s}$, for any $k=0,1,2,...,e-1$, and $ k^{'}=0,1,2,...,2e-1$.\\
(II) If $f$ is odd, then
$$x^{3lp^{s}}-\lambda=\widehat{B}_{0}(ax)^{p^{s}}\widehat{B}_{l}(ax)^{p^{s}}\prod_{k=0}^{e-1}\widehat{B}_{g^{k}}(ax)^{p^{s}}\widehat{B}_{3g^{k}}(ax)^{p^{s}},$$
  where $0\leq k\leq e-1$.

 Therefore, we have
 $$C=\langle\widehat{B}_{0}(ax)^{\varepsilon}\widehat{B}_{l}(ax)^{\rho}\prod_{k=0}^{e-1}\widehat{B}_{g^{k}}(ax)^{\tau_{k}}\widehat{B}_{3g^{k}}(ax)^{\upsilon_{k}}\rangle,$$
 $$C^{\perp}=\langle\widehat{B}_{0}(a^{-1}x)^{p^{s}-\varepsilon}\widehat{B}_{-l}(a^{-1}x)^{p^{s}-\rho}\prod_{k=0}^{e-1}\widehat{B}_{-g^{k}}(a^{-1}x)^{p^{s}-\tau_{k}}\widehat{B}_{-3g^{k}}(a^{-1}x)^{p^{s}-\upsilon_{k}}\rangle,$$
  where $0\leq \varepsilon,\rho,\tau_{k},\upsilon_{k}\leq p^{s}$, for any $k=0,1,2,...,e-1$.\\

  \noindent\textbf{Proof.} From Theorems 2.2 and 3.3, we see that this theorem is obtained immediately.\qed
\dse{4.2~~All constacyclic codes of length $3lp^{s}$ over $F_{q}$ when $d=3$}
In this subsection, we consider the second case, i.e., $d=\textrm{gcd}(q-1,3lp^{s})=3$. In this case, we have that $F_{q}^{*}=\langle\xi\rangle=\langle\xi^{3}\rangle\bigcup\xi^{p^{s}}\langle\xi^{3}\rangle\bigcup\xi^{2p^{s}}\langle\xi^{3}\rangle$ is the decomposition of cosets of $F_{q}^{*}$ over subgroup $\langle\xi^{3lp^{s}}\rangle$ by Lemma 3.1. Therefore, it is enough to consider the cyclic codes, $\xi^{p^{s}}-$constcyclic codes and $\xi^{2p^{s}}-$constcyclic codes if we want to determine all constacyclic codes of length $3lp^{s}$ over $F_{q}$, when gcd$(q-1,3lp^{s})=3$.

From the Section 2, we have that the irreducible factorization of $x^{l}-1$ is given by $x^{l}-1=\prod_{i=0}^{e}M_{i}(x),$ where $M_{i}(x)=\prod_{j\in C_{i}}(x-\eta^{i})$ with $\eta$ is a primitive $l$th root of unity. According to this, we deduce the following lemma, which gives the irreducible factorization of $x^{lp^{s}}-\xi^{j}$ in $F_{q}[x]$, when $\textrm{gcd}(q-1,l)=1$.\\

\noindent\textbf{Lemma 4.2.} Let gcd$(q-1,l)=1$. Then, for any $\xi^{j}\in F_{q}^{*}$, there exists a unique element $b_{j}\in F_{q}^{*}$ such that $b_{j}^{lp^{s}}\xi^{j}=1$, where $j=0,1,2,...q-1$. Thus, the irreducible factorization of $x^{lp^{s}}-\xi^{j}$ is given by $$x^{lp^{s}}-\xi^{j}=\prod_{i=0}^{e}\widehat{M}_{i}(b_{j}x)^{p^{s}},$$
 where $M_{i}(x)$ is the minimal polynomial of the $q-$cyclotomic coset $C_{i}$ modulo $l$. \\

\noindent\textbf{Proof.} Similar to Lemma 3.1, we have the decomposition of coset of $F_{q}^{*}=\langle\xi\rangle$ over the subgroup $\langle\xi^{lp^{s}}\rangle$ is given by $F_{q}^{*}=\langle\xi\rangle$, when $\textrm{gcd}(q-1,l)=1$. Next, by Theorem 3.3, we have the result.\qed\\

\noindent\textbf{Lemma 4.3.} Let $\textrm{gcd}(q-1,3lp^{s})=3$. Then the irreducible factorization of $x^{3lp^{s}}-1$ is given by
$$x^{3lp^{s}}-1=\prod_{i=0}^{e}M_{i}(x)^{p^{s}}\widehat{M}_{i}(b_{\frac{q-1}{3}}x)^{p^{s}}\widehat{M}_{i}(b_{\frac{2(q-1)}{3}}x)^{p^{s}},$$
where $M_{i}(x)$ is the minimal polynomial of the $q-$cyclotomic coset $C_{i}$ modulo $l$ and $b_{\frac{q-1}{3}}$, $b_{\frac{2(q-1)}{3}}\in F_{q}^{*}$.

\noindent\textbf{Proof.} As $\textrm{gcd}(q-1,3lp^{s})=3$, then it is clear that $\xi^{\frac{q-1}{3}}\in F_{q}^{*}$ is a primitive $3$th root of unity. Hence, we have $$x^{3lp^{s}}-1=(x^{l}-1)^{p^{s}}(x^{l}-\xi^{\frac{q-1}{3}})^{p^{s}}(x^{l}-\xi^{\frac{2(q-1)}{3}})^{p^{s}}.$$

Further, by Lemma 4.2, we get
$$x^{3lp^{s}}-1=\prod_{i=0}^{e}M_{i}(x)^{p^{s}}\widehat{M}_{i}(b_{\frac{q-1}{3}}x)^{p^{s}}\widehat{M}_{i}(b_{\frac{2(q-1)}{3}}x)^{p^{s}},$$
where $M_{i}(x)$ is the minimal polynomial of the $q-$cyclotomic coset $C_{i}$ modulo $l$ and $b_{\frac{q-1}{3}}$, $b_{\frac{2(q-1)}{3}}\in F_{q}^{*}$.\qed\\

Before determining $\xi^{ip^{s}}-$constacyclic codes, $i=1,2$, we must explicitly decompose the polynomial $x^{3lp^{s}}-\xi^{ip^{s}}$ $(i=1,2)$, into the product of monic irreducible factors. Obviously, we only need to determine the irreducible factorization of $x^{3l}-\xi^{i}$ $(i=1,2)$. Firstly, we consider the polynomial $x^{3}-\xi^{i}$ $(i=1,2)$.
Since, $x^{3}-\xi^{i}$ $(i=1,2)$ is irreducible in $F_{q}[x]$, we get that $F_{q^{3}}$ is a splitting field for  $x^{3}-\xi^{i}$ $(i=1,2)$ over $F_{q}$. Thus, there exists $\nu_{i}\in F_{q^{3}}$ such that $\nu_{i}^{3}=\xi^{i}$ $(i=1,2)$. Further, it is easy to get that $\nu_{i},\alpha\nu_{i}$, and $\alpha^{2}\nu_{i}$ are all the roots of $x^{3}-\xi^{i}$ $(i=1,2)$, in $F_{q^{3}}$, where $\alpha$ is a primitive $3$th root of unity. In addition, we see that $\nu_{i}\in F_{q^{3}}$ but $\nu_{i} \notin F_{q}$, and $\nu_{i}$ is primitive $3(q-1)$th root of unity, $i=1,2$.

As $\textrm{gcd}(3(q-1),l)=1$, we can find a bijection $\theta$ from the set $D$ to itself such that $\theta(\nu)=\nu^{l}$, for any $\nu\in D$, where $D$ consists of all the primitive $3(q-1)th$ roots of unity of $F_{q^{3}}^{*}$. Therefore, there exists a unique element $\omega_{i}\in D$ such that $\nu_{i}^{-1}=\theta(\omega_{i})=\omega_{i}^{l}$, i.e., $\omega_{i}^{l}\nu_{i}=1$, $i=1,2$.

From the above discussion, we have the following two lemmas.\\

\noindent\textbf{Lemma 4.4.} The irreducible factorization of $x^{3l}-\xi$ over $F_{q}$ is given as follows:\\
 (I) If $\textrm{gcd}(f,3)=1$,
        $$x^{3l}-\xi=\prod_{i=0}^{e}R_{i}(x),$$
where $R_{i}(x)=\widehat{M}_{i}(\omega_{1}x)\widehat{M}_{i}(\alpha\omega_{1}x)\widehat{M}_{i}(\alpha^{2}\omega_{1}x)$ for any $i=0,1,2,...,e$, with $\omega_{1}$ is a primitive $3(q-1)$th root of unity and $\alpha$ is a primitive $3$th root of unity.\\
(II) If $\textrm{gcd}(f,3)=3$,

 $$x^{3l}-\xi=P(x)\prod_{i=1}^{e}Q_{i}(x)U_{i}(x)Z_{i}(x),$$
where
$P(x)=(x-\omega_{1}^{-1})(x-\alpha\omega_{1}^{-1})(x-\alpha^{2}\omega_{1}^{-1})$, $Q_{i}(x)=\widehat{A}_{i}(\omega_{1}x)\widehat{A}_{iq}(\alpha\omega_{1}x)\widehat{A}_{iq^{2}}(\alpha^{2}\omega_{1}x)$,
$U_{i}(x)=\widehat{A}_{i}(\alpha\omega_{1}x)\widehat{A}_{iq}(\alpha^{2}\omega_{1}x)\widehat{A}_{iq^{2}}(\omega_{1}x)$,
and
$Z_{i}(x)=\widehat{A}_{i}(\alpha^{2}\omega_{1}x)\widehat{A}_{iq}(\omega_{1}x)\widehat{A}_{iq^{2}}(\alpha\omega_{1}x)$ for any $i=1,2,...,e$, with $\omega_{1}$ is a primitive $3(q-1)$th root of unity, $\alpha$ is a primitive $3$th root of unity.\\

 \noindent\textbf{Proof.} (I) If$\textrm{gcd}(f,3)=1$, we get that $C_{0}$ and $C_{k}$,$1\leq k\leq e=\frac{\phi(l)}{f}$ are all the $q^{3}-$cyclotomic cosets modulo $l$. Let $\nu_{1}$ be a root of $x^{3}-\xi$, and $\alpha$ be a primitive $3$th root of unity. Then we have $\nu_{1},\alpha\nu_{1}$, and $\alpha^{2}\nu_{1}$ are all the roots of $x^{3}-\xi$ over $F_{q^{3}}$, i.e., $$ x^{3l}-\xi =(x^{l}-\nu_{1})(x^{l}-\alpha\nu_{1})(x^{l}-\alpha^{2}\nu_{1}).$$
 From the above discussion, we know that there exists $\omega_{1}$ such that $\omega_{1}^{l}\nu_{1}=1$, where $\omega_{1}$ is a primitive $3(q-1)$th root of unity. Hence, we have $\omega_{1}^{q}=\alpha\omega_{1}$ or $\alpha^{2}\omega_{1}$.
 Next, we just consider $\omega_{1}^{q}=\alpha\omega_{1}$. However, for the case $\omega_{1}^{q}=\alpha^{2}\omega_{1}$, we see that the result is still right in the same way. As $\textrm{gcd}(3,l)=1$, we know $l\equiv1(\textrm{mod}~3)$ or $l\equiv2(\textrm{mod}~3)$. When $l\equiv2(\textrm{mod}~3)$, we have $(\alpha\omega_{1})^{l}\alpha\nu_{1}=\alpha^{l+1}=1$ and $(\alpha^{2}\omega_{1})^{l}\alpha^{2}\nu_{1}=\alpha^{2(l+1)}=1$. When $l\equiv1(\textrm{mod}~3)$, we have  $(\alpha^{2}\omega_{1})^{l}\alpha\nu_{1}=\alpha^{2l+1}=1$ and  $(\alpha\omega_{1})^{l}\alpha^{2}\nu_{1}=\alpha^{l+2}=1$. Hence, there always exist $\omega_{1},\alpha\omega_{1}$ and $\alpha^{2}\omega_{1}$ such that $\omega_{1}^{l}\nu_{1}=1$, $(\alpha\omega_{1})^{l}\alpha\nu_{1}=1$ and $(\alpha^{2}\omega_{1})^{l}\alpha^{2}\nu_{1}=1$ or $\omega_{1}^{l}\nu_{1}=1$, $(\alpha^{2}\omega_{1})^{l}\alpha\nu_{1}=1$ and $(\alpha\omega_{1})^{l}\alpha^{2}\nu_{1}=1$.
 Further, by Lemma 4.2, we get $$x^{3l}-\xi=\prod_{i=0}^{e}\widehat{M}_{i}(\omega_{1}x)\widehat{M}_{i}(\alpha\omega_{1}x)\widehat{M}_{i}(\alpha^{2}\omega_{1}x),$$
which is the monic irreducible factorization of $x^{3l}-\xi$ over $F_{q^{3}}$. And we have $\widehat{M}_{i}(\omega_{1}x)=\prod_{k\in C_{i}}(x-\omega_{1}^{-1}\eta^{k})$, $\widehat{M}_{i}(\alpha\omega_{1}x)=\prod_{k\in C_{i}}(x-\alpha^{2}\omega_{1}^{-1}\eta^{k})$ and $\widehat{M}_{i}(\alpha^{2}\omega_{1}x)=\prod_{k\in C_{i}}(x-\alpha\omega_{1}^{-1}\eta^{k})$.
Obviously, when $k$ runs over $C_{i}$, $\omega_{1}^{-1}\eta^{k}$ gives all the roots of $\widehat{M}_{i}(\omega_{1}x)$.
As $\omega_{1}^{q}=\alpha\omega_{1}$ and $kq,kq^{2}\in C_{i}$, we have $(\omega_{1}^{-1}\eta^{k})^{q}=\alpha^{2}\omega_{1}^{-1}\eta^{kq}$ and $(\omega_{1}^{-1}\eta^{k})^{q^{2}}=\alpha\omega_{1}^{-1}\eta^{kq^{2}}$, which gives a root of  $\widehat{M}_{i}(\alpha\omega_{1}x)$ and $\widehat{M}_{i}(\alpha^{2}\omega_{1}x)$ respectively. Therefore, it's easy to deduce that $\widehat{M}_{i}(\omega_{1}x)\widehat{M}_{i}(\alpha^{2}\omega_{1}x) \\\widehat{M}_{i}(\alpha^{2}\omega_{1}x)$
 is a irreducible polynomial over $F_{q}$.\\
 (II) When $\textrm{gcd}(f,3)=3$, we have that $A_{0},A_{k},A_{kq}, A_{kq^{2}}$ consist of all the distinct $q^{3}-$cyclotomic cosets modulo $l$, where $1\leq k\leq e$. Then, the irreducible factorization of $x^{l}-1$ over $F_{q^{3}}$ is given by$$x^{l}-1=A_{0}(x)A_{1}(x)A_{q}(x)A_{q^{2}}(x)A_{2}(x)A_{2q}(x)A_{2q^{2}}(x)...A_{e}(x)A_{eq}(x)A_{eq^{2}}(x).$$
 Next, in the same way as (I), we can proved the conclusion (II) holds.\qed\\

 Using arguments similar to the proof in Lemma 4.3, we have the following lemma, and we omit its proof here.\\

 \noindent\textbf{Lemma 4.5.}  The irreducible factorization of $x^{3l}-\xi^{2}$ over $F_{q}$ is given as follows:\\
 (I) If $\textrm{gcd}(f,3)=1$,
       $$ x^{3l}-\xi^{2} =\prod_{i=0}^{e}R_{i}^{'}(x),$$
where $R_{i}^{'}(x)=\widehat{M}_{i}(\omega_{2}x)\widehat{M}_{i}(\alpha\omega_{2}x)\widehat{M}_{i}(\alpha^{2}\omega_{2}x)$ for any $i=0,1,2,...,e$, with $\omega_{2}$ is a primitive $3(q-1)$th root of unity and $\alpha$ is a primitive $3$th root of unity.\\
(II) If $\textrm{gcd}(f,3)=3$,

        $$x^{3l}-\xi^{2}=P^{'}(x)\prod_{i=1}^{e}Q_{i}^{'}(x)U_{i}^{'}(x)Z_{i}^{'}(x),$$
where $P^{'}(x)=(x-\omega_{2}^{-1})(x-\alpha\omega_{2}^{-1})(x-\alpha^{2}\omega_{2}^{-1})$, $Q^{'}_{i}(x)=\widehat{A}_{i}(\omega_{2}x)\widehat{A}_{iq}(\alpha\omega_{2}x)\widehat{A}_{iq^{2}}(\alpha^{2}\omega_{2}x)$,
$U^{'}_{i}(x)=\widehat{A}_{i}(\alpha\omega_{2}x)\widehat{A}_{iq}(\alpha^{2}\omega_{2}x)\widehat{A}_{iq^{2}}(\omega_{2}x)$, and $Z^{'}_{i}(x)=\widehat{A}_{i}(\alpha^{2}\omega_{2}x)\widehat{A}_{iq}(\omega_{2}x)
\widehat{A}_{iq^{2}}(\alpha\omega_{2}x)$ for any $i=1,2,...,e$, with $\omega_{2}$ is a primitive $3(q-1)$th root of unity and $\alpha$ is a primitive $3$th root of unity.\\

Combining Lemmas 4.3, 4.4 and 4.5, we easily obtain the next result.\\

\noindent\textbf{Theorem 4.6.} Let $\textrm{gcd}(q-1,3lp^{s})=3$ and $\alpha=\xi^{\frac{q-1}{3}}$. For any element $\lambda$ of $F_{q}^{*}$ and $\lambda-$constacyclic code $C$ of length $3lp^{s}$ over $F_{q}$, one of the following cases holds:\\
(I) If $\lambda\in\langle\xi^{3}\rangle$, then there exists some element $c\in F_{q}^{*}$ such that $c^{3lp^{s}}\lambda=1$, and we have
$$C=\langle\prod_{i=0}^{e}\widehat{M}_{i}(cx)^{\varepsilon_{i}}\widehat{M}_{i}(cb_{\frac{q-1}{3}}x)^{\sigma_{i}}\widehat{M}_{i}(cb_{\frac{2(q-1)}{3}}x)^{\tau_{i}}\rangle,$$
 $$C^{\perp}=\langle\prod_{i=0}^{e}\widehat{M}_{-i}(c^{-1}x)^{p^{s}-\varepsilon_{i}}\widehat{M}_{-i}(c^{-1}b^{-1}_{\frac{q-1}{3}}x)^{p^{s}-\sigma_{i}}\widehat{M}_{-i}(c^{-1}b^{-1}_{\frac{2(q-1)}{3}}x)^{p^{s}-\tau_{i}}\rangle,$$
where $0\leq\varepsilon_{i},\sigma_{i},\tau_{i}\leq p^{s}$ for any $i=0,1,2,...,e$.\\
(II) If $\lambda\in\xi^{p^{s}}\langle\xi^{3}\rangle$, then there exists some element $c_{1}\in F_{q}^{*}$ such that $c_{1}^{3lp^{s}}\lambda=\xi^{p^{s}}$, and one of the following holds:\\
$~~~~$(i) If $\textrm{gcd}(f,3)=1$,
$$C=\langle\prod_{i=0}^{e}\widehat{R}_{i}(c_{1}x)^{\varepsilon}\rangle,$$
$$C^{\perp}=\langle\prod_{i=0}^{e}\widehat{R}_{-i}(c_{1}^{-1}x)^{p^{s}-\varepsilon}\rangle,$$
where $0\leq\varepsilon\leq p^{s}$, $R_{-i}(x)=\widehat{M}_{-i}(\omega^{-1}_{1}x)\widehat{M}_{-i}(\alpha^{2}\omega^{-1}_{1}x)\widehat{M}_{-i}(\alpha\omega^{-1}_{1}x)$ for any $i=0,1,2,...,e$.\\
$~~~~$(ii) If $\textrm{gcd}(f,3)=3$,
$$C=\langle \widehat{P}(c_{1}x)^{\varepsilon}\prod_{i=1}^{e}\widehat{Q}_{i}(c_{1}x)^{\varepsilon_{i}}\widehat{U}_{i}(c_{1}x)^{\sigma_{i}}\widehat{Z}_{i}(c_{1}x)^{\tau_{i}}\rangle,$$
$$C^{\perp}=\langle \widehat{P}^{*}(c^{-1}_{1}x)^{p^{s}-\varepsilon}\prod_{i=1}^{e}\widehat{Q}_{-i}(c^{-1}_{1}x)^{p^{s}-\varepsilon_{i}}\widehat{U}_{-i}(c^{-1}_{1}x)^{p^{s}-\sigma_{i}}\widehat{Z}_{-i}(c^{-1}_{1}x)^{p^{s}-\tau_{i}}\rangle,$$
where $0\leq \varepsilon_{i},\sigma_{i},\tau_{i}\leq p^{s}$,
$P^{*}(x)=(x-\omega_{1})(x-\alpha^{2}\omega_{1})(x-\alpha\omega_{1})$, $Q_{-i}(x)=\widehat{A}_{-i}(\omega^{-1}_{1}x)\cdot\widehat{A}_{-iq}(\alpha^{2}\omega^{-1}_{1}x)\widehat{A}_{-iq^{2}}(\alpha\omega^{-1}_{1}x)$,
$U_{-i}(x)=\widehat{A}_{-i}(\alpha^{2}\omega^{-1}_{1}x)\widehat{A}_{-iq}(\alpha\omega^{-1}_{1}x)\widehat{A}_{-iq^{2}}(\omega^{-1}_{1}x)$,
and
$Z_{-i}(x)=\widehat{A}_{-i}(\alpha\omega^{-1}_{1}x)\widehat{A}_{-iq}(\omega^{-1}_{1}x)\widehat{A}_{-iq^{2}}(\alpha^{2}\omega^{-1}_{1}x)$, for any $i=1,2,...,e$.\\
(III) If $\lambda\in\xi^{2p^{s}}\langle\xi^{3}\rangle$, then there exists some element $c_{2}\in F_{q}^{*}$ such that $c_{2}^{3lp^{s}}\lambda=\xi^{2p^{s}}$, and one of the following holds:\\
 $~~~~$(i) If $\textrm{gcd}(f,3)=1$,
 $$C=\langle\prod_{i=0}^{e}\widehat{R^{'}}_{i}(c_{2}x)^{\varepsilon_{i}}\rangle,$$
 $$C^{\perp}=\langle\prod_{i=0}^{e}\widehat{R^{'}}_{-i}(c^{-1}_{2}x)^{p^{s}-\varepsilon_{i}}\rangle,$$
where $0\leq\varepsilon\leq p^{s}$, $R_{-i}^{'}(x)=\widehat{M}_{-i}(\omega^{-1}_{2}x)\widehat{M}_{-i}(\alpha^{2}\omega^{-1}_{2}x)\widehat{M}_{-i}(\alpha\omega^{-1}_{2}x)$ for any $i=0,1,2,...,e$.\\
$~~~~$(ii) If $\textrm{gcd}(f,3)=3$,
$$C=\langle \widehat{P^{'}}(c_{2}x)^{\varepsilon}\prod_{i=1}^{e}\widehat{Q^{'}}_{i}(c_{2}x)^{\varepsilon_{i}}\widehat{U^{'}}_{i}(c_{2}x)^{\sigma_{i}}\widehat{Z^{'}}_{i}(c_{2}x)^{\tau_{i}})\rangle,$$
$$C^{\perp}=\langle \widehat{P^{'}}^{*}(c^{-1}_{2}x)^{p^{s}-\varepsilon}\prod_{i=1}^{e}\widehat{Q^{'}}_{-i}(c^{-1}_{2}x)^{p^{s}-\varepsilon_{i}}\widehat{U^{'}}_{-i}(c^{-1}_{2}x)^{p^{s}-\sigma_{i}}\widehat{Z^{'}}_{-i}(c^{-1}_{2}x)^{p^{s}-\tau_{i}})\rangle,$$
where $0\leq \varepsilon_{i},\sigma_{i},\tau_{i}\leq p^{s}$,
$P^{'*}(x)=(x-\omega_{2})(x-\alpha^{2}\omega_{2})(x-\alpha\omega_{2})$, $Q^{'}_{-i}(x)=\widehat{A}_{-i}(\omega_{2}^{-1}x)\cdot\widehat{A}_{-iq}(\alpha^{2}\omega^{-1}_{2}x)\widehat{A}_{-iq^{2}}(\alpha\omega^{-1}_{2}x)$,
$U^{'}_{-i}(x)=\widehat{A}_{-i}(\alpha^{2}\omega^{-1}_{2}x)\widehat{A}_{-iq}(\alpha\omega^{-1}_{2}x)\widehat{A}_{-iq^{2}}(\omega^{-1}_{2}x)$,
and
$Z^{'}_{-i}(x)=\widehat{A}_{-i}(\alpha\omega^{-1}_{2}x)\widehat{A}_{-iq}(\omega^{-1}_{2}x)\widehat{A}_{-iq^{2}}(\alpha^{2}\omega^{-1}_{2}x)$ for any $i=1,2,...,e$.\\

\dse{4.3~~All constacyclic codes of length $3lp^{s}$ over $F_{q}$ when $d=l$}
Let $d=\textrm{gcd}(q-1,3lp^{s})=l$, i.e., $l|(q-1)$ and $\textrm{gcd}(q-1,3)=1$. Then $\eta=\xi^{\frac{q-1}{l}}\in F_{q}^{*}$ is a primitive $l$th root of unity. Therefore, we have the following lemma.\\

\noindent\textbf{Lemma 4.7.} Assume that $\textrm{gcd}(q-1,3lp^{s})=l$, and let $\eta=\xi^{\frac{q-1}{l}}$. Then, the irreducible factorization of $x^{3lp^{s}}-1$ over $F_{q}$ is given by:
$$x^{3lp^{s}}-1=\prod_{k=0}^{l-1}(x-\eta^{k})^{p^{s}}(x^{2}+\eta^{k}x+\eta^{2k})^{p^{s}}.$$\\

\noindent\textbf{Proof.} As $\textrm{gcd}(q-1,3)=1$, there exists no primitive $3$th root of unity in $F_{q}$, which implies $x^{2}+x+1$ is irreducible. By this, we can deduce that $x^{2}+\eta^{k}x+\eta^{2k}$ is irreducible, for any $k=1,2,...,l$. Otherwise $x^{2}+\eta^{k}x+\eta^{2k}$ is reducible, then we have $\eta^{-2k}(x^{2}+\eta^{k}x+\eta^{2k})=(\eta^{-k} x)^{2}+\eta^{-k}x+1$ is reducible. By substituting $x$ for $\eta^{-k}x$ in the above polynomial, we have that $x^{2}+x+1$ is reducible, which is a contradiction. Since $\eta=\xi^{\frac{q-1}{l}}$ be a primitive $l$th root of unity, then $\eta^{3}$ is also a primitive $l$th root of unity. Hence, the irreducible factorization of $x^{3lp^{s}}-1$ over $F_{q}$ is given by
$$x^{3lp^{s}}-1=(x^{3l}-1)^{p^{s}}=\prod_{k=0}^{l-1}(x^{3}-\eta^{3k})^{p^{s}}=\prod_{k=0}^{l-1}(x-\eta^{k})^{p^{s}}(x^{2}+\eta^{k}x+\eta^{2k})^{p^{s}}.$$\qed\\

\noindent\textbf{Lemma 4.8.} Assume that $\textrm{gcd}(q-1,3lp^{s})=l$, then the irreducible factorization of $x^{3lp^{s}}-\xi^{jp^{s}}$, $1\leq j\leq l-1$, over $F_{q}$ is given by:\\
(I) When $\textrm{gcd}(3,j)=3$, let $j=3k$, for some integer $k$. Then we have
$$x^{3lp^{s}}-\xi^{jp^{s}}=(x^{l}-\xi^{k})^{p^{s}}(x^{2l}+\xi^{k}x^{l}+\xi^{2k})^{p^{s}}.$$
(II) When $\textrm{gcd}(3,j)=1$, there must exist some integer $i$, $1\leq i\leq q-1$, such that $3i=j+q-1$ or $3i=j+2(q-1)$. Then we have
        $$x^{3lp^{s}}-\xi^{jp^{s}}=(x^{l}-\xi^{i})^{p^{s}}(x^{2l}+\xi^{i}x^{l}+\xi^{2i})^{p^{s}}.$$

\noindent\textbf{Proof.} (I) Obviously, $\textrm{gcd}(l,k)=1$. From Lemma 2.3, it is very easy to verify that $x^{l}-\xi^{k}$ is irreducible. By the proof of Lemma 4.5, we see that $x^{2}+\xi^{k}x+\xi^{2k}$ is irreducible over $F_{q}$.
Now, we suppose that  $\delta$ is any root of $x^{2}+\xi^{k}x+\xi^{2k}$ in some extended field of $F_{q}$, and $e$ is the order of $\delta$. Then, we have $\delta^{3}=\xi^{3k}$. Further, we deduce that $\frac{e}{(e,3)}=\frac{q-1}{(q-1,3k)}$, i.e., $\frac{e}{(e,3)}=\frac{q-1}{(q-1,k)}$, as $\textrm{gcd}(q-1,3)=1$. By the reduction again, we get $e=\frac{(q-1)(e,3)}{(q-1,k)}$. From Lemma 2.4, we can verify that $x^{2l}+\xi^{k}x^{l}+\xi^{2k}$ is irreducible.\\
$~~~~$(II) When $(3,j)=1$, we get that $x^{3}-\xi^{j}$, $1\leq j\leq l-1$, are all reducible, from Lemma 2.3. Therefore, there must exist some $\xi^{i}\in F_{q}^{*}$, which is a root of  $x^{3}-\xi^{j}$, for any $j=1,2,...l-1$. Then, $\xi^{3i}-\xi^{j}=0$, i.e., $\xi^{3i}=\xi^{j}$.
As $1\leq i\leq q-1$ and $1\leq j\leq l-1$, we deduce that $3i=j+q-1$ or $3i=j+2(q-1)$ and $\textrm{gcd}(l,i)=1$.
Next, working similar to the proof of (I), we get that the results (i) and (ii) hold.\qed\\

According to Lemma 4.7 and Lemma 4.8, we get the following theorem immediately.\\

\noindent\textbf{Theorem 4.9.}  Assume that $\textrm{gcd}(q-1,3lp^{s})=l$, and let $\eta=\xi^{\frac{q-1}{l}}$. For any element $\lambda$ of $F_{q}^{*}$ and $\lambda-$constacyclic code $C$ of length $3lp^{s}$ over $F_{q}$, one of the following cases holds:\\
(I) If $\lambda\in\langle\xi^{l}\rangle$, then there exists $c_{1}\in F_{q}^{*}$ such that $c_{1}^{3lp^{s}}\lambda=1$, and we have\\
$$C=\langle\prod_{k=0}^{l-1}(x-c_{1}^{-1}\eta^{k})^{\varepsilon_{k}}(x^{2}+c_{1}^{-1}\eta^{k}x+c_{1}^{-2}\eta^{2k})^{\tau_{k}}\rangle,$$
$$C^{\perp}=\langle\prod_{k=0}^{l-1}(x-c_{1}\eta^{-k})^{p^{s}-\varepsilon_{k}}(x^{2}+c_{1}\eta^{-k}x+c_{1}^{2}\eta^{-2k})^{p^{s}-\tau_{k}}\rangle,$$
where $0\leq \varepsilon_{k},\tau_{k}\leq p^{s}$, for any $k=0,1,2,...,l-1$.\\
(II) If $ \lambda\in\xi^{jp^{s}}\langle\xi^{l}\rangle$, $1\leq j\leq l-1$, then there exists $c_{2}\in F_{q}^{*}$ such that $c_{2}^{3lp^{s}}\lambda=\xi^{jp^{s}}$, and one of the following holds:\\
$~~~~$(i) When $(3,j)=3$, let $j=3k$, for some integer $k$. Then we have
$$C=\langle(x^{l}-c_{2}^{-l}\xi^{k})^{\varepsilon_{k}}(x^{2l}+c_{2}^{-l}\xi^{k}x^{l}+c_{2}^{-2l}\xi^{2k})^{\tau_{k}}\rangle,$$
$$C^{\perp}=\langle(x^{l}-c_{2}^{l}\xi^{-k})^{p^{s}-\varepsilon_{k}}(x^{2l}+c_{2}^{l}\xi^{-k}x^{l}+c_{2}^{2l}\xi^{-2k})^{p^{s}-\tau_{k}}\rangle,$$
where $0\leq\varepsilon_{k},\tau_{k}\leq p^{s}$.\\
$~~~~$(ii) When $\textrm{gcd}(3,j)=1$, there must exist some integer $i$, $1\leq i\leq q-1$, such that $3i=j+q-1$ or $3i=j+2(q-1)$. Then we have
$$C=\langle(x^{l}-c_{2}^{-l}\xi^{i})^{\varepsilon_{i}}(x^{2l}+c_{2}^{-l}\xi^{i}x^{l}+c_{2}^{-2l}\xi^{2i})^{\tau_{i}}\rangle,$$
$$C^{\perp}=\langle(x^{l}-c_{2}^{l}\xi^{-i})^{p^{s}-\varepsilon_{i}}(x^{2l}+c_{2}^{l}\xi^{-i}x^{l}+c_{2}^{2l}\xi^{-2i})^{p^{s}-\tau_{i}}\rangle.$$
where $0\leq \varepsilon_{i},\tau_{i}\leq p^{s}$.\\
\dse{4.4~~All constacyclic codes of length $3lp^{s}$ over $F_{q}$ when $d=3l$}
In this subsection, we assume that $d=\textrm{gcd}(3lp^{s},q-1)=3l$, namely $3l|(q-1)$. Clearly, there exists an element $\gamma=\xi^{\frac{q-1}{3l}}\in F_{q}^{*}$, which is a primitive $3l$th root of unity. Further, due to $l|(q-1)$, and $3|(q-1)$, it is easy to know that $\eta=\xi^{\frac{q-1}{l}}$ and $\beta=\xi^{\frac{q-1}{3}}$ are primitive $l$th and $3$th roots of unity respectively. \\

From Lemma 3.1, we get that the $F_{q}^{*}=\langle\xi\rangle=\langle\xi^{3l}\rangle\cup\xi^{p^{s}}\langle\xi^{3l}\rangle\cup\xi^{2p^{s}}\langle\xi^{3l}\rangle\cup...\cup\xi^{(3l-1)p^{s}}\langle\xi^{3l}\rangle$.
Therefore, any element $\lambda$ of $F_{q}^{*}$ belongs to exactly one of the cosets, i.e., there is a unique integer $j$, $0\leq j\leq 3l-1$, such that $\lambda\in \xi^{jp^{s}}\langle\xi^{3l}\rangle$, namely $\lambda-$constacyclic codes are equivalent to $\xi^{jp^{s}}-$constacyclic codes. Hence, we just need to determine $\xi^{jp^{s}}-$constacyclic codes, where $0\leq j\leq 3l-1$.\\

\noindent\textbf{Lemma 4.10.} Let $d=\textrm{gcd}(3lp^{s},q-1)=3l$ and $\gamma=\xi^{\frac{q-1}{3l}}$. Then irreducible factorization of $x^{3lp^{s}}-1$ over $F_{q}$ is given as follows:
$$x^{3lp^{s}}-1=\prod_{i=0}^{3l-1}(x-\gamma^{i})^{p^{s}}.$$\\

\noindent\textbf{Lemma 4.11.} Let $\eta=\xi^{\frac{q-1}{l}}$ and $\beta=\xi^{\frac{q-1}{3}}$. Then the irreducible factorization of $x^{3lp^{s}}-\xi^{jp^{s}}$ over $F_{q}$ is given as follows:\\
(I) When $\textrm{gcd}(3l,j)=l$, we have
$$x^{3lp^{s}}-\xi^{jp^{s}}=(x^{3l}-\xi^{tl})^{p^{s}}=\prod_{i=0}^{l-1}(x^{3}-\xi^{t}\eta^{i})^{p^{s}},$$
where $t=1$ or $2$.\\
(II) When $\textrm{gcd}(3l,j)=3$, we have
$$x^{3lp^{s}}-\xi^{jp^{s}}=(x^{3l}-\xi^{3k})^{p^{s}}=\prod_{i=0}^{2}(x^{l}-\xi^{k}\beta^{i})^{p^{s}},$$
where $k$ is some integer such that $j=3k$.\\
(III) Otherwise, we can see that $\textrm{gcd}(3l,j)=1$. Then we have
$$x^{3lp^{s}}-\xi^{jp^{s}}=(x^{3l}-\xi^{j})^{p^{s}}.$$\\

\noindent\textbf{Proof.} $(I)$ As $\textrm{gcd}(3l,j)=l$ and $1\leq j\leq 3l-1$, we have $j=tl$, where $t=1,2$. Obviously, $\eta=\xi^{\frac{q-1}{l}}$ is a primitive $l-$th root of unity in $F_{q}$. Therefore, we get $$x^{3lp^{s}}-\xi^{jp^{s}}=(x^{3l}-\xi^{tl})^{p^{s}}=\prod_{i=0}^{l-1}(x^{3}-\xi^{t}\eta^{i})^{p^{s}},$$
Next, we prove that the polynomial $x^{3}-\xi^{t}\eta^{i}$, for any $i=0,1,2,...,l-1$, is irreducible in $F_{q}[x]$.
Firstly, we know that the multiplicative order of $\xi^{t}\eta^{i}=\xi^{t+\frac{i(q-1)}{l}}$, $t=1,2$, is $e_{i}=\frac{q-1}{(q-1,t+\frac{i(q-1)}{l})}$. As $3|(q-1)$, $\textrm{gcd}(3,t)=1$ and $\textrm{gcd}(3,l)=1$, we get that $(3,t+\frac{i(q-1)}{l})=1$.
Thus, $3$ divides $e_{i}$ but not $\frac{q-1}{e_{i}}=(q-1,t+\frac{i(q-1)}{l})$. From Lemma 2.3, we get the polynomial $x^{3}-\xi^{t}\eta^{i}$, for any $i=0,1,2,...,l-1$, is irreducible in $F_{q}[x]$. In the same way, we have (II) and (III) hold.\qed\\

In the following theorem, we determine all constacyclic codes of length $3lp^{s}$ over $F_{q}$ and their dual codes, when $d=\textrm{gcd}(3lp^{s},q-1)=3l$.\\

\noindent\textbf{Theorem 4.12.} Assume that $\textrm{gcd}(3lp^{s},q-1)=3l$, let $\gamma=\xi^{\frac{q-1}{3l}},\eta=\xi^{\frac{q-1}{l}}$ and $\beta=\xi^{\frac{q-1}{3}}$ be primitive $3l$th, $l$th and $3$th root of unity in $F_{q}$ respectively. For any element $\lambda$ of $F_{q}^{*}$ and $\lambda-$ constcyclic codes $C$ of length $3lp^{s}$ over $F_{q}$. One of the following holds:\\
(I) If $\lambda\in\langle\xi^{3l}\rangle$, then there exists $d_{1}\in F_{q}^{*}$ such that $d_{1}^{3lp^{s}}\lambda=1$, and we have
$$C=\langle\prod_{i=0}^{3l-1}(x-d_{1}^{-1}\gamma^{i})^{\varepsilon_{i}}\rangle,$$
$$C^{\perp}=\langle\prod_{i=0}^{3l-1}(x-d_{1}\gamma^{-i})^{p^{s}-\varepsilon_{i}}\rangle,$$ where $0\leq\varepsilon_{i}\leq p^{s}$, for any $i=0,1,2,...,3l-1$.\\
(II) If $\lambda\in\xi^{jp^{s}}\langle\xi^{3l}\rangle$, $1\leq j\leq 3l-1$, then there exists $d_{2}\in F_{q}^{*}$ such that $d_{2}^{3lp^{s}}\lambda=\xi^{jp^{s}}$, and one of the following holds:\\
$~~~~$(i) When $\textrm{gcd}(3l,j)=l$, we have
$$C=\langle\prod_{i=0}^{l-1}(x^{3}-d_{2}^{-3}\xi^{t}\eta^{i})^{\varepsilon_{i}}\rangle,$$
$$C^{\perp}=\langle\prod_{i=0}^{l-1}(x^{3}-d_{2}^{3}\xi^{-t}\eta^{-i})^{p^{s}-\varepsilon_{i}}\rangle,$$
where $0\leq \varepsilon_{i}\leq p^{s}$, for any $i=0,1,2,...,l-1$.\\
$~~~~$(ii) When $\textrm{gcd}(3l,j)=3$, we have
$$C=\langle\prod_{i=0}^{2}(x^{l}-d_{2}^{-l}\xi^{k}\beta^{i})^{\varepsilon_{i}}\rangle,$$
$$C^{\perp}=\langle\prod_{i=0}^{2}(x^{l}-d_{2}^{l}\xi^{-k}\beta^{-i})^{p^{s}-\varepsilon_{i}}\rangle,$$
where $0\leq \varepsilon_{i}\leq p^{s}$ for $i=0,1,2$, and $j=3k$.\\
$~~~~$(iii) When $\textrm{gcd}(3l,j)=1$, we have
$$C=\langle (x^{3l}-d_{2}^{-3l}\xi^{j})^{\varepsilon}\rangle,$$
$$C^{\perp}=\langle (x^{3l}-d_{2}^{3l}\xi^{-j})^{p^{s}-\varepsilon}\rangle,$$
where $0\leq\varepsilon\leq p^{s}$.

\dse{5~~All self-dual cyclic codes of length $3lp^{s}$ over $F_{q}$}
In Section 4, we gave the generator polynomials of all the constacyclic codes and their dual codes of length $3lp^{s}$ over $F_{q}$. We will determine all the self-dual cyclic codes of length $3lp^{s}$ over $F_{q}$ in detail, in this section.

It is well known that there exist self-dual cyclic codes of length $N$ over $F_{q}$ if and only if $N$ is even and the characteristic of $F_{q}$ is $p=2$ $[10,11]$. Therefore, we get that self-dual cyclic codes of length $3lp^{s}$ over $F_{q}$ exist only when $p=2$.

Let $x^{3lp^{s}}-1=(x^{3l}-1)^{p^{s}}=f_{1}(x)^{p^{s}}f_{2}(x)^{p^{s}}\cdot\cdot\cdot f_{a}(x)^{p^{s}}h_{1}(x)^{p^{s}}h^{*}_{1}(x)^{p^{s}}\cdot\cdot\cdot h_{b}(x)^{p^{s}}h^{*}_{b}(x)^{p^{s}}$ be the irreducible factorization of $x^{3lp^{s}}-1$, where $f_{1}(x),f_{2}(x),...,f_{a}(x)$ are monic irreducible self-reciprocal polynomials over $F_{q}$, $h_{j}(x)$ and its reciprocal polynomial $h^{*}_{j}(x)$, $1\leq j\leq b$, are also monic irreducible polynomials over $F_{q}$. Hence, for any cyclic code $C=\langle g(x)\rangle$ of length $3lp^{s}$ over $F_{q}$, we suppose that $$g(x)=f_{1}(x)^{\tau_{1}}f_{2}(x)^{\tau_{2}}\cdot\cdot\cdot f_{a}(x)^{\tau_{a}}h_{1}(x)^{\delta_{1}}h^{*}_{1}(x)^{\sigma_{1}}\cdot\cdot\cdot h_{b}(x)^{\delta_{b}}h^{*}_{b}(x)^{\sigma_{b}},$$ where $0\leq \tau_{i},\delta_{j},\sigma_{j}\leq p^{s}$, for any $i=1,2,...,a$, and $j=1,2,...,b$. Then, we have $$h(x)=f_{1}(x)^{p^{s}-\tau_{1}}f_{2}(x)^{p^{s}-\tau_{2}}\cdot\cdot\cdot f_{a}(x)^{p^{s}-\tau_{a}}h_{1}(x)^{p^{s}-\delta_{1}}h^{*}_{1}(x)^{p^{s}-\sigma_{1}}\cdot\cdot\cdot h_{b}(x)^{p^{s}-\delta_{b}}h^{*}_{b}(x)^{p^{s}-\sigma_{b}}.$$ Therefore,
$$h^{*}(x)=f_{1}(x)^{p^{s}-\tau_{1}}f_{2}(x)^{p^{s}-\tau_{2}}\cdot\cdot\cdot f_{a}(x)^{p^{s}-\tau_{a}}h_{1}(x)^{p^{s}-\sigma_{1}}h^{*}_{1}(x)^{p^{s}-\delta_{1}}\cdot\cdot\cdot h_{b}(x)^{p^{s}-\sigma_{b}}h^{*}_{b}(x)^{p^{s}-\delta_{b}}.$$

If $C$ is a self-dual cyclic code, we get the following theorem.\\

\noindent\textbf{Theorem 5.1.} With the above notations, we have that $C$ is a self-dual cyclic code if and only if $2\tau_{i}=p^{s}$, $1\leq i\leq a$, and $\delta_{j}+\sigma_{j}=p^{s}$, $1\leq j\leq b$.\\

\noindent\textbf{Proof.} $C$ is a self-dual cyclic code if and only if $g(x)=h^{*}(x)$, i.e., $2\tau_{i}=p^{s}$, $1\leq i\leq a$, and $\delta_{j}+\sigma_{j}=p^{s}$, $1\leq j\leq b$.\qed\\

According to this theorem, we see that it is enough to determine the irreducible factorization of $x^{3lp^{s}}-1$ as above. And if we do this, we can give all the self-dual cyclic codes immediately.

Similar to the definition of reciprocal polynomial, we give the following definition.\\

\noindent\textbf{Definition 5.2.} Let $C_{s}=\{s,sq,...,sq^{f-1}\}$ be any $q-$cyclotomic coset modulo $l$, then $$C_{s}^{*}=\{-s,-sq,...,-sq^{f-1}\}$$
is said to be the reciprocal coset of $C_{s}$. The coset $C_{s}$ is called self-reciprocal if $C_{s}=C_{s}^{*}$.\\

Obviously, $C^{*}_{s}$ is still a $q-$cyclotomic coset modulo $l$. And the reciprocal polynomial of the minimal polynomial of $C_{s}$ is the minimal polynomial of $C^{*}_{s}$. Hence, the minimal polynomial of $C_{s}$ is also self-reciprocal if $C_{s}$ is self-reciprocal.\\

\noindent\textbf{Lemma 5.3.} Assume that $q\equiv1(\textrm{mod}~3)$. For the $q-$cyclotomic cosets, which have been described in Lemma 2.1, one of the following holds:\\
(I) If $f=\textrm{ord}_{l}(q)$ is even, we have
$$B_{0}^{*}=B_{0},B_{l}^{*}=B_{-l},B_{g^{k}}^{*}=B_{-g^{k}},B_{3g^{k}}^{*}=B_{3g^{k}},$$
where $0\leq k\leq e-1$.\\
(II) If $f=\textrm{ord}_{l}(q)$ is odd, we have
$$B_{0}^{*}=B_{0},B_{l}^{*}=B_{-l},B_{g^{k}}^{*}=B_{-g^{k}},B_{3g^{k^{'}}}^{*}=B_{-3g^{k^{'}}}$$
where $\{B_{3g^{k}}\}=\{B_{3g^{k^{'}}}\}\bigcup\{B_{-3g^{k^{'}}}\}$ and $0\leq k\leq e-1$, $0\leq k^{'}\leq \frac{e}{2}-1$.\\

\noindent\textbf{Proof.} (I) Obviously, we only need to prove $B_{3g^{k}}^{*}=B_{3g^{k}}$. If $f=\textrm{ord}_{l}(q)$ is even, we deduce that $q^{\frac{f}{2}}\equiv -1(\textrm{mod}~l)$. According to this, we get there exist $i,j$, $0\leq i,j\leq f-1$, and $|j-i|=\frac{f}{2}$, such that $3g^{k}q^{i}\equiv -3g^{k}q^{j}(\textrm{mod}~3l)$, for any $3g^{k}q^{i}\in B_{3g^{k}}$, $0\leq k\leq e-1$. Therefore, we have $B_{3g^{k}}^{*}=B_{3g^{k}}$, for any $k=0,1,...,e-1$.\\
$~~~~$(II) In the same way as Lemma 2.1, we get that $B_{0},B_{l},B_{-l},B_{g^{k}},B_{-g^{k}},B_{3g^{k^{'}}}$ and $B_{-3g^{k^{'}}}$, $0\leq k\leq e-1$, $0\leq k^{'}\leq \frac{e}{2}-1$, are all the distinct $q-$cyclotomic cosets modulo $3l$. The next result is obvious. \qed\\

\noindent\textbf{Lemma 5.4.} Let $q\equiv2(\textrm{mod}~3)$ and $f$ be even. For the $q-$cyclotomic cosets, which have been described in Lemma 2.1, one of the following holds:\\
(I) When $f=2t$ and $t$ is even, we have $$B_{0}^{*}=B_{0},B_{l}^{*}=B_{l},B_{g^{k}}^{*}=B_{-g^{k}},B_{3g^{k}}^{*}=B_{3g^{k}},$$
where $\{B_{g^{k^{'}}}\}=\{B_{g^{k}}\}\bigcup \{B_{-g^{k}}\}$,$0\leq k\leq e-1$ and $0\leq k^{'}\leq 2e-1$.\\
(II) When $f=2t$ and $t$ is odd, we have $$B_{0}^{*}=B_{0},B_{l}^{*}=B_{l},B_{g^{k^{'}}}^{*}=B_{g^{k^{'}}},B_{3g^{k}}^{*}=B_{3g^{k}},$$
where $0\leq k\leq e-1$ and $0\leq k^{'}\leq 2e-1$. \\

\noindent\textbf{Proof.} (I) In the same way as Lemma 2.1, we get $B_{0},B_{l},B_{g^{k}},B_{-g^{k}}$ and $B_{3g^{k}}$ , $0\leq k\leq e-1$, are all the distinct $q-$cyclotomic cosets modulo $3l$. Next, We first prove that $B_{l}^{*}=B_{l}$, i.e., $\{l,lq\}^{*}=\{l,lq\}$. As $q\equiv2(\textrm{mod}~3)$, i.e., $q\equiv-1(\textrm{mod}~3)$, then $lq\equiv-l(\textrm{mod}~3)$. Since $\textrm{gcd}(3,l)=1$, we have $lq\equiv-l(\textrm{mod~}3l)$, which implies $B_{l}^{*}=B_{l}$. Otherwise, by the conclusion (I) of Lemma 5.3, we get the  other results immediately.\\
$~~~~$(II) According to (I), it is obvious that we only need to prove $B_{g^{k^{'}}}^{*}=B_{g^{k^{'}}}$. As $t$ is odd, we have
$q^{t}\equiv-1(\textrm{mod}~3)$. Since $q^{t}\equiv-1(\textrm{mod}~l)$ and $\textrm{gcd}(3,l)=1$, we get $q^{t}\equiv-1(\textrm{mod}~3l)$. Then, we deduce that there exist $i,j$, $0\leq i,j\leq f-1$, and $|j-i|=t$, such that $g^{k^{'}}q^{i}\equiv -g^{k^{'}}q^{j}(\textrm{mod}~3l)$, for any $g^{k^{'}}q^{i}\in B_{k^{'}}$, $0\leq k^{'}\leq 2e-1$, i.e., $B_{g^{k^{'}}}^{*}=B_{g^{k^{'}}}$. \qed\\

\noindent\textbf{Lemma 5.5.} Let $q\equiv2(\textrm{mod}~3)$ and $f$ be odd. For the $q-$cyclotomic cosets, which have been described in Lemma 2.1, we have
$$B_{0}^{*}=B_{0},B_{l}^{*}=B_{l},B_{g^{k^{'}}}^{*}=B_{-g^{k^{'}}},B_{3g^{k^{'}}}^{*}=B_{-3g^{k^{'}}},$$
where $\{B_{g^{k}}\}=\{B_{g^{k^{'}}}\}\bigcup \{B_{-g^{k^{'}}}\}$, $\{B_{3g^{k}}\}=\{B_{3g^{k^{'}}}\}\bigcup\{B_{-3g^{k^{'}}}\}$ and $0\leq k\leq e-1$, $0\leq k^{'}\leq \frac{e}{2}-1$.\\

From the above lemmas, we can give all the self-dual cyclic codes of length $3\cdot2^{s}l$ over $F_{2^{m}}$ and their enumeration in the following theorem.\\

\noindent\textbf{Theorem 5.6.} Let $l\neq 3$ be an odd prime, $p=2$, $f=\textrm{ord}_{l}(2^{m})$, and $e=\frac{l-1}{f}$. Then, for cyclic self-dual codes of length $3\cdot2^{s}l$ over $F_{2^{m}}$, we have \\
(I) When $q\equiv 1(\textrm{mod}~3)$, one of the following hold:\\
$~~~~$(i) If $f=\textrm{ord}_{l}(q)$ is even, then there exist $(2^{s}+1)^{e+1}$ cyclic self-dual codes of length $3\cdot2^{s}l$ over $F_{2^{m}}$ given by $$\langle(x-1)^{2^{s-1}}B_{l}(x)^{\delta}B_{-l}(x)^{2^{s}-\delta}\prod_{k=0}^{e-1}B_{g^{k}}(x)^{\delta_{k}}B_{-g^{k}}(x)^{2^{s}-\delta_{k}}B_{3g^{k}}(x)^{2^{s-1}}\rangle,$$
where $0\leq\delta ,\delta_{k}\leq 2^{s}$, for any $0\leq k\leq e-1$.\\
$~~~~$(ii) If $f=\textrm{ord}_{l}(q)$ is odd, then there exist $(2^{s}+1)^{\frac{3e}{2}+1}$ cyclic self-dual codes of length $3\cdot2^{s}l$ over $F_{2^{m}}$ given by $$\langle(x-1)^{2^{s-1}}B_{l}(x)^{\delta}B_{-l}(x)^{2^{s}-\delta}\prod_{k=0}^{e-1}B_{g^{k}}(x)^{\delta_{k}}B_{-g^{k}}(x)^{2^{s}-\delta_{k}}\prod_{k^{'}=0}^{\frac{e}{2}-1}B_{3g^{k^{'}}}(x)^{\sigma_{k^{'}}}B_{-3g^{k^{'}}}(x)^{2^{s}-\sigma_{k^{'}}}\rangle,$$
where $0\leq \delta, \delta_{k},\sigma_{k^{'}}\leq 2^{s}$, for any $0\leq k\leq e$ and $0\leq k^{'}\leq \frac{e}{2}-1$.\\
(II) When $q\equiv 2(\textrm{mod}~3)$, we have\\
$~~~~$(i) If $f=2t$ and $t$ is even, then there exist $(2^{s}+1)^{e}$ cyclic self-dual codes of length $3\cdot2^{s}l$ over $F_{2^{m}}$ given by $$\langle(x-1)^{2^{s-1}}B_{l}(x)^{2^{s-1}}\prod_{k=0}^{e-1}B_{g^{k}}(x)^{\delta_{k}}B_{-g^{k}}(x)^{2^{s}-\delta_{k}}B_{3g^{k}}(x)^{2^{s-1}}\rangle,$$
where $0\leq \delta_{k}\leq 2^{s}$, for any $0\leq k\leq e-1$.\\
$~~~~$(ii) If $f=2t$ and $t$ is odd, then there exists only one cyclic self-dual code of length $3\cdot2^{s}l$ over $F_{2^{m}}$ given by $$\langle(x-1)^{2^{s-1}}B_{l}(x)^{2^{s-1}}\prod_{k^{'}=0}^{2e-1}B_{g^{k^{'}}}(x)^{2^{s-1}}\prod_{k=0}^{e-1}B_{3g^{k}}(x)^{2^{s-1}}\rangle.$$
$~~~~$(iii) If $f$ is odd, then there exist $(2^{s}+1)^{e}$ cyclic self-dual codes of length $3\cdot2^{s}l$ over $F_{2^{m}}$ given by $$\langle(x-1)^{2^{s-1}}B_{l}(x)^{2^{s-1}}\prod_{k^{'}=0}^{\frac{e}{2}-1}B_{g^{k^{'}}}(x)^{\delta_{k^{'}}}B_{-g^{k^{'}}}(x)^{2^{s}-\delta_{k^{'}}}B_{3g^{k^{'}}}(x)^{\sigma_{k^{'}}}B_{-3g^{k^{'}}}(x)^{2^{s}-\sigma_{k^{'}}}\rangle.$$
where $0\leq \delta_{k^{'}},\sigma_{k^{'}}\leq 2^{s}$, for any $0\leq k^{'}\leq \frac{e}{2}-1$.

\end{document}